\documentclass[twocolumn]{aastex63}
\usepackage{natbib}
\usepackage{needspace}
\usepackage{amsmath}

\graphicspath{{./}{figures/}}
\newcommand\Modot{$\rm M_{\odot}$}
\newcommand\Msun{$\rm M_{\odot}$}
\newcommand\Rodot{$\rm R_{\odot}$}
\newcommand\MJup{$\rm M_{J}$}
\newcommand\RJup{$\rm R_{J}$}
\newcommand\MEarth{$\rm M_{\oplus}$}
\newcommand\Lsun{$\rm L_{\odot}$}

\newcommand\Li{$^7$Li}
\newcommand\Lisix{$^6$Li}

\newcommand\Benine{$^9\mathrm{Be}$}
\newcommand\Be{$^7\mathrm{Be}$}
\newcommand\ALi{$A(\mathrm{Li})$}
\newcommand\planetmass{1.0~\MJup}

\newcommand\maxdist{$R_{\star}\sim~0.1~\mathrm{AU}~=~22$~\Rodot{}}
\newcommand\logg{log($g$)}

\shortauthors{Soares-Furtado et al.}
\begin{document}
\title{Lithium Enrichment Signatures of Planetary Engulfment Events in Evolved Stars}

\correspondingauthor{Melinda Soares-Furtado}
\email{mmsoares@wisc.edu}

\author[0000-0001-7493-7419]{M. Soares-Furtado}
\altaffiliation{National Science Foundation Graduate Research Fellow}
\affiliation{Department of Astronomy,  University of Wisconsin-Madison, 475 N. Charter St., Madison, WI 53703, USA}
\affiliation{Department of Astrophysical Sciences, Princeton University, Princeton, NJ 08544, USA}

\author[0000-0002-8171-8596]{Matteo Cantiello}
\affiliation{Department of Astrophysical Sciences, Princeton University, Princeton, NJ 08544, USA}
\affiliation{Center for Computational Astrophysics, Flatiron Institute, 162 5th Avenue, New York, NY 10010,
USA}

\author[0000-0002-1417-8024]{Morgan MacLeod}
%\altaffiliation{NASA Einstein Fellow}
\affiliation{Harvard-Smithsonian Center for Astrophysics, 60 Garden Street, Cambridge, MA, 02138, USA}

\author[0000-0001-5082-6693]{Melissa K.~Ness}
\affiliation{Center for Computational Astrophysics, Flatiron Institute, 162 5th Avenue, New York, NY 10010,
USA}
\affiliation{Department of Astronomy, Columbia University, Pupin Physics Laboratories, New York, NY 10027, USA}

%%%%%%%%%%%%%%%%%%%%%%%%%%%%%%%%%%%%%%%%%%%%%%%%%%%%%%%%%%%%%%%%%%%%%%%%%%%%%%%%
\begin{abstract}
Planetary engulfment events have long been proposed as a lithium (Li) enrichment mechanism contributing to the population of Li-rich giants (\ALi{}~$\geq  1.5$~dex).
Using MESA stellar models and \ALi{} abundance measurements obtained by the GALAH survey, we calculate the strength and observability of the surface Li enrichment signature produced by the engulfment of a hot Jupiter (HJ). 
We consider solar-metallicity stars in the mass range of $1-2$~\Msun{} and the Li supplied by a HJ of \planetmass{}. 
We explore engulfment events that occur near the main sequence turn-off (MSTO) and out to orbital separations of \maxdist{}.
We map our results onto the Hertzsprung-Russell Diagram, revealing the statistical significance and survival time of lithium enrichment.
We identify the parameter space of masses and evolutionary phases where the engulfment of a HJ can lead to Li enrichment signatures at a $5\sigma$ confidence level and with meteoritic abundance strengths.
The most compelling strengths and survival times of engulfment-derived Li enrichment are found among host stars of 1.4~\Msun{} near the MSTO.
Our calculations indicate that planetary engulfment is not a viable enrichment pathway for stars that have evolved beyond the subgiant branch. For these sources, observed Li enhancements are likely to be produced by other mechanisms, such as the Cameron--Fowler process or the accretion of material from an asymptotic giant branch (AGB) companion.
Our results do not account for second-order effects, such as extra mixing processes, which can further dilute Li enrichment signatures.
\end{abstract}

%%%%%%%%%%%%%%%%%%%%%%%%%%%%%%%%%%%%%%%%%%%%%%%%%%%%%%%%%%%%%%%%%%%%%%%%%%%%%%%%
\section{Introduction} 
\label{sec:intro}
%Introduction

%Li isotopes
Lithium (Li) is a trace element found in two stable isotopes, \Li{} and \Lisix{}.
%, with cosmic abundances of $^7\mathrm{Li}/\mathrm{H}~=10^{-9}$ and $^6\mathrm{Li}/\mathrm{H} = 10^{-10}$ \citep{Zappala1972}.  %$^7\mathrm{Li}/\mathrm{H}~=~9.2~\times~10^{-10}$ and $^6\mathrm{Li}/\mathrm{H} = 7.6\times~10^{-11}$ \citep{Zappala1972}. 
An isotope ratio of \Li{}/\Lisix{} $\sim 12$ is largely ubiquitous among presolar grains \citep[e.g.,][]{Lyon2007,King2012}, meteorites \citep[e.g.,][]{Chaussidon1998,Chaussidon1999,Sephton2004}, planets \citep[e.g.,][]{Chan2000,Tomascak2004}, of F-, G-, and K-type dwarf stars \citep[e.g.,][]{Soderblom1985}, and local star-forming regions \citep[e.g.,][]{Knauth2003}.
While observations reveal some spatial Li abundance variation within the interstellar medium (ISM) \citep{Knauth2003}, the consistently dominant isotope is \Li{}.
Given its preponderance, our work focuses on the engulfment signatures of the \Li{} isotope.

%Li destruction in pMS
The \Li{} isotope is destroyed via proton fusion ($\mathrm{p}~+~^7\mathrm{Li}~\rightarrow~\mathrm{^4He}~+~\mathrm{^4He}$)  at temperatures $\gtrsim 2.5~\times~10^6$~K \citep{Bodenheimer1965}.
Given that the \Li{}-burning temperature is $8\times$ less than the hydrogen ignition temperature, \Li{} depletion commences during the pre-main sequence (pMS) phase  \citep[e.g.,][]{Bodenheimer1965,Deliyannis1990,Piau2002}.
The chemical constituents of a fully-convective pMS star are well mixed, thereby aiding in the efficacy of \Li{} depletion.
Observations of solar metallicity field stars that have settled onto the main sequence (MS) reveal depletion from an initial meteoritic strength of \ALi{}~$\sim 3.3$~dex \citep[e.g.,][]{Grevesse1989,Bildsten1997,Asplund2009} to \ALi{}~$\sim2.4$~dex \citep[e.g.,][]{Carlos2019}. 
%We provide a description of the 12-point scale used to calculate these abundance measurements in Appendix~\ref{sec:appendixltt}.

Steady \Li{} depletion has been observed among  solar metallicity MS stars between $1-2$~\Msun{}  \citep[e.g.,][]{Deliyannis2000,Baumann2010,Monroe2013,Melendez2014,Carlos2016,Carlos2019}. 
Standard stellar evolutionary models do not account for MS depletion, nor do they account for the observed dispersion of \ALi{} measurements.
As stars evolve away from the MS, they undergo the first dredge-up (FDU) process, whereby the inner boundary of the expanding convective envelope begins to overlap with regions containing \Li{}-depleted, H-processed material.
This further dilutes the \Li{} surface abundance.
Standard stellar evolutionary models estimate abundances of \ALi{}~${\sim}1.5$~dex for $1.0$~\Msun{}stars at this evolutionary phase \citep{Iben1967,Iben1967b}.
This is supported by GALAH photospheric measurements, which indicate a nominal \Li{} abundance of \ALi{}~$\sim 1.5$~dex among G--K field giants at early phases of the RGB.
The surface abundance drops to \ALi{}~${\sim}~0.5$~dex for stars near the RGB luminosity bump (LB).

Observations of \Li{}-rich giant stars are therefore puzzling. 
One pathway to \Li{} enrichment is the engulfment or accretion of a close-orbiting substellar companion, as substellar objects do not achieve the requisite internal temperatures to deplete their initial meteoritic \Li{} supply \citep[e.g.,][]{Alexander1967,Siess1999,Israelian2002}. 
This enrichment mechanism would be rather commonplace, as $30\%$ of sun-like stars have been found to host substellar companions with orbital periods $<400$~days \citep{Zhu2018}. 
Additionally, $1\%$ of sun-like stars are found to host a hot Jupiter (HJ) companion---gas giants with orbital periods $\lesssim 10$~days and orbital separations $\lesssim 0.1$~AU \citep{Winn2015}.

Dynamic in-spiral timescales are on the order of few years among red giant branch (RGB) hosts, rendering it unlikely to observe a planetary engulfment event \citep{Staff2016}.
Estimates of planetary engulfment occurrence rates are on the order of $0.1-1$ event per year in the Galaxy \citep{MacLeod2018}.
More plausible is the detection of long-lasting signatures produced by planetary engulfment events, such as chemical enrichment \citep[e.g.,][]{Adamow2012}.

\textit{Can the engulfment of a HJ companion induce statistically significant, long-standing \Li{} enrichment signatures among stellar hosts?}
To answer this question, we investigate the strength and duration of the \Li{} enrichment signature produced by the ingestion and subsequent compositional mixing of a HJ.
We examine stars between $1-2$~\Msun{} and a \planetmass{} companion ($9.55 \times 10^{-4}$ \Msun{}).
We investigate systems across a wide range of orbital separations ($a$), exploring engulfment when $R_{\star}=a$ out to a maximum separation of \maxdist{}. 

We investigate host stars that offer promising opportunities for observational follow-up.
Stars $<1$~\Msun{} are more unlikely to have evolved beyond the MS, while stars $>2$~\Msun{} undergo short subgiant phases, reducing the likelihood of detection at this stage. 
We employ MESA (Modules for Experiments in Stellar Astrophysics) stellar evolutionary models to calculate the strength of the \Li{} enrichment signature at different evolutionary phases  \citep{Paxton2011,Paxton2013,Paxton2015,Paxton2018}. 
Photospheric \ALi{} surface abundance measurements obtained from the GALAH (GALactic Archaeology with HERMES) survey allow us to calculate the survival time and statistical significance of the enrichment signatures \citep{Buder2018}. 
In Section~\ref{sec:Lithium}, we describe the theory of stellar lithium abundances for stars on the MS, subgiant, and giant branches. 
Additionally, we review proposed lithium depletion and enrichment mechanisms. 
In Section~\ref{sec:GALAH}, we discuss our calculation of the stellar \ALi{} baseline and the associated variance. 
In Section~\ref{sec:Engulfment}, we discuss our MESA models.
We present our predicted engulfment-derived \ALi{} enrichment signatures and their expected survival time.
In Section~\ref{sec:Discussion}, we identify systems where statistically significant engulfment-derived enrichment signatures are expected to arise.
We review the criterion required for the total dissolution of an engulfed substellar companion.
We then present several intriguing planetary engulfment candidates revealed by observational techniques.
In Section~\ref{sec:summary}, we summarize our findings and discuss future endeavors.

%%%%%%%%%%%%%%%%%%%%%%%%%%%%%%%%%%%%%%%%%%%%%%%%%%%%%%%%%%%%%%%%%%%%%%%%%%%%%%%%
%\needspace{3\baselineskip}
%\section{Theory of Stellar Surface Lithium Abundances}
\section{Surface Lithium Abundances}
\label{sec:Lithium}
%In Section~\ref{subsec:MSdepletion}, we discuss \Li{} depletion mechanisms at various phases of stellar evolution among stars of $1-2$~\Msun{}.
%We provide a short summary of the \Li{} abundance among lower mass stars in Appendix~\ref{sec:appendixlowmass}.
%In Section~\ref{subsec:enrichedgiants}, we present observations of \Li{}-enriched giant stars.
%In  Section~\ref{subsec:Limechanisms} we discuss viable \Li{} enrichment mechanisms.
%.............................................................
%Li destruction on the MS
\subsection{Lithium Depletion Mechanisms}
\label{subsec:MSdepletion}
%Li in 1-2 solar mass stars 
%During the MS phase of $1-2$~\Msun{} stars, the convective base temperature does not reach the \Li{}-burning threshold.
Standard stellar evolutionary models---which omit rotation, convective overshoot, diffusion, and mass loss---predict a relatively fixed \ALi{} abundance during the MS phase of $1-2$~\Msun{} stars \citep[e.g.,][]{Pinsonneault1997}. 
Yet, observations of MS stars reveal ongoing \Li{} depletion  \citep[e.g.,][]{Baumann2010,Monroe2013,Melendez2014,Carlos2016,Carlos2019}.
This suggests that physical mechanisms omitted from standard stellar models may be responsible for \Li{} depletion during this phase \citep[see for review][]{Michaud1991}.

This includes the mechanism driving the Li-dip---a strong Li depletion feature observed among $1-1.5$~\Msun{} stars with effective temperatures of $6400-6800$~K \citep[e.g.,][]{Boesgaard1986,Hobbs1988}. 
The Li-dip is prominent among stellar clusters older than $100$~Myr.
While a complete picture of the mechanisms driving the Li-dip is lacking, rotation-induced mixing has been identified as an important component \citep[e.g.,][]{Boesgaard1987,Deliyannis1998}.
%The correlated depletion of lithium and beryllium suggests that mass-loss and diffusion are not the drivers of this effect \citep{Deliyannis1998}.

Several proposed depletion mechanisms share the theme of mixing \Li{}-depleted, H-processed material from the depths of the star into the convective envelope.
These mechanisms include mixing induced by overshooting during the pMS phase \citep{Fu2015,Thevenin2017},
mixing induced by overshooting during the MS phase \citep[e.g.,][]{Bohm1963,Ahrens1992,Xiong2009}, mixing induced by gravity waves \citep[e.g.,][]{Press1981,Montalban1996}, rotation-induced turbulent diffusion \citep[e.g.,][]{Baglin1985,Zhang2012}, and differential rotation \citep[e.g.,][]{Bouvier1995,Brun1999}.

Mixing can also be impacted by the transfer of angular momentum from exterior reservoirs, such as an accretion disk \citep{Bouvier1993,Bouvier1995} or binary companion \citep[e.g.,][]{Zahn1994,Beck2017}.
Some studies correlate \ALi{} abundance strengths to the presence of substellar companions \citep[e.g.,][]{King1997,Israelian2004,Israelian2009,Delgado2014}.
\cite{Adamow2018} reported an increased likelihood of planets around \Li{}-rich giants, while \cite{Carlos2019} speculated that specific solar system architectures could play an important role in stellar \ALi{} abundances.  
Yet, tension remains regarding  the  role  and  viability  of  such  mechanisms.
For example, \cite{Baumann2010} found no such connection between \ALi{} abundances and the presence of planets.

%As described in Section~\ref{sec:intro}, the FDU process significantly dilutes the surface abundance of \Li{}, reducing \ALi{} by a factor of $30-60$, depending on stellar mass and metallicity  \citep{Iben1967,Iben1967b}. 
%\footnote{The LB---also known as the ``red giant bump"---occurs when the outward-moving H-burning shell reaches the base of the convective envelope. As the H-burning shell enters a chemically homogeneous region, it crosses a composition discontinuity, resulting in a non-monotonic change in the stellar luminosity \citep[e.g.,][]{FusiPecci1990}.}

Nonstandard mixing processes such as thermohaline mixing may play an important role in altering the \Li{} abundance among post-MS stars.
Thermohaline mixing arises from an inversion of the mean molecular weight created by $^3\mathrm{He}$-burning outside the H-burning shell, and is an important mechanism among low-mass stars near the LB as well as intermediate-mass stars on the early-asymptotic giant branch (AGB) phase \citep[e.g.,][]{Charbonnel2007,Charbonnel2010,Cantiello2010,Lattanzio2015}.
While thermohaline mixing is found to contribute to \Li{} destruction, the numerical schemes and assumptions employed in stellar evolutionary codes result in \ALi{} abundances that differ by many orders of magnitude  \citep{Lattanzio2015}. 
Other nonstandard mixing processes among giant stars include magneto-thermohaline mixing \citep{Denissenkov2009} and mixing induced by magnetic buoyancy \citep[e.g.,][]{Busso2007,Nordhaus2008}. 
%.............................................................
\subsection{Lithium Enriched Giants} 
\label{subsec:enrichedgiants}
%Stars enriched in \Li{} have been observed within a wide range of environments, including the Galactic disk  \citep[e.g.,][]{Balachandran2000, Monaco2011}, the Galactic halo \citep[e.g.,][]{Deepak2019}, the Galactic bulge \citep[e.g.,][]{Gonzalez2009}, among isolated and multiple stars in open clusters \citep[e.g.,][]{Anthony-Twarog2013,Monaco2014,Pasquini2014,Anthony-Twarog2018}, within metal-poor and metal-rich globular clusters \citep[e.g.,][]{Ruchti2011, DOrazi2015,Kirby2016}, and within dwarf galaxies \citep[e.g.,][]{Kirby2012}. 
Stars enriched in \Li{} have been observed across all phases of evolution \citep[e.g.,][]{Monaco2011,Lebzelter2012,Martell2013}.
Most puzzling are the detections of \Li{}-rich giants (\ALi{}~$\gtrsim 1.5$~dex), which comprise ${\sim}1$\% of the G--K giant population \citep[e.g.,][]{Wallerstein1982,Brown1989,Charbonnel2010,Deepak2019,Gao2019}.
%\footnote{Observations of 1.8~\Msun{} stars in Trumpler 20 indicate that this occurrence rate may be as large as $5\%$ for more massive stars \citep{Aguilera2016b}.} 
%The first detection of a Li-rich giant was observed by \cite{}.
To date, more than 10,000 Li-rich evolved stars have been observed \citep[e.g.,][]{Brown1989,Charbonnel2000,Balachandran2000,Reddy2005,Carlberg2010,Kumar2011,Martell2013,Adamow2014,Adamow2015,Yan2018,Li2018,Deepak2019,Zhou2019,Casey2019,Singh2019,Gao2019}.
%Enrichment may if investigations employed observationally-derived \ALi{} variance measurements to discern a statistically-significant threshold for stars of specified masses, ages, and chemical compositions---as opposed to the conventional \ALi{}~$\geq 1.5$~dex enrichment threshold.

Perhaps most remarkable is the population of super-rich giants, which harbor surface abundance measurements exceeding meteoritic levels (\ALi{}~$\geq 3.3$~dex).
The most \Li{}-rich giant observed to date was found to have a surface abundance of \ALi{}~$\sim 4.9$~dex \citep{Gao2019}. 
Super-rich giants, which make up $6\%$ of the enriched giant population \citep[e.g.,][]{Balachandran2000,Zhou2019,Singh2019}, cannot be explained by the preservation of the initial stellar \Li{} supply.

%.............................................................
\subsection{Lithium  Enrichment Mechanisms}
\label{subsec:Limechanisms}

%One spatially ubiquitous enrichment mechanism is cosmic ray spallation, which is driven by interactions between high-energy cosmic rays and stellar carbon/oxygen atoms  \citep{Mitler1972}.
%This process has been found to contribute no more than 10\% of the excess \Li{} supply in stars and is unable to reproduce nominal Li isotope ratios---cosmic ray spallation results in \Li{}/\Lisix{}~$\sim 1.4$, while the observed isotope ratio is \Li{}/\Lisix{}~$ \sim 12$ \citep{Reeves1994}. 
In this section, we review viable \Li{} enrichment mechanisms for stars of $1-2$~\Msun{} at evolutionary phases extending from the MS to the RGB.
These mechanisms can be grouped into three categories: (a) self-generation, (b) mass transfer, and (c) the engulfment of a substellar companion.

Proponents of self-generated enrichment mechanisms have proposed two critical evolutionary stages, during which the star experiences intrinsic \Li{} enrichment that is detectable within a stellar photosphere.
These stages include the LB \citep[e.g.,][]{Charbonnel2000} and the helium-core burning phase at the red clump region \citep[e.g.,][]{Kumar2011,Monaco2014,Deepak2019,Singh2019}.
It is important to note that there are significant challenges in the identification of the precise evolutionary state of a star. 
As a result, \Li{}-enriched sources can be misclassified due to the ambiguity associated with the ${T_\mathrm{eff}}-L$ plane of the Hertzsprung-Russell (H-R) diagram. 

Some research teams have begun to incorporate asteroseismic age estimates along with their spectroscopic analysis to mitigate ambiguity in the stellar evolutionary state.
Unfortunately, this has led to mixed results. 
For example, an asteroseismic analysis was conducted on 24 super Li-rich giants discovered in the LAMOST survey data  \citep{Singh2019}. 
%None of the enriched sources in this sample were located along the RGB.
They found that the enriched sources were preferentially found within the red clump region. 
Additionally, super-rich sources are preferentially found in the red clump region.
The pathway to \Li{} enrichment for red clump stars has been connected to the merger of a helium (He) white dwarf and an RGB stellar companion \citep{Zhang2020}. 
%An analysis of the rotation rates of giant stars in clusters performed by \cite{Carlberg2014} also found that fast rotators were preferentially found within the red clump region.

In contrast, \cite{Casey2019} unveiled \Li{}-rich 2,330 giant stars between $1-3$~\Msun{}.
While asteroseismological analysis reveals that 80\% of these enriched giants were undergoing the AGB phase, sources were also found at the LB and RGB tip \citep{Casey2019}.
The team concluded that enrichment could not be purely attributed to single star evolution, indicating that accretion from AGB companions or substellar companions may play an important role.

\subsubsection{Cameron-Fowler Mechanism}
The Cameron-Fowler (CF) mechanism provides a viable pathway to surface \Li{} enhancement. To operate, \Be{} from the PP-II chain of H-burning must be rapidly transported to cooler regions of a convective envelope, where the isotope is converted into \Li{} via $\beta$-decay \citep{Cameron1971}. Rapid mixing is required for this mechanism to function, as the short \Be{} decay time must exceed the mixing timescale.
%\footnote{The \Be{} half-life is generally ${\sim}53$~days, however, \cite{Cameron1955} found that this can be extended by the ionization state of the isotope.}

The existence of the CF mechanism is strongly supported by observational results showing \Li{}-rich AGB stars to be more prevalent than \Li{}-rich RGB stars \citep[e.g.,][]{Smith1989,Plez1993,Deepak2019,Singh2019}. 
A survey performed on 25 globular clusters found \Li{} enrichment among 0.2\% of the RGB stars and 1.6\% of the AGB stars \citep{Kirby2016}.
Indeed, the convective envelopes of AGB stars reach depths where the temperatures and densities are sufficient for the CF mechanism to operate.  

Observational data indicate that the CF mechanism is in operation after stars reach the LB \citep[e.g.,][]{Gilroy:1991,Gratton2000},
when the mean molecular weight barrier is erased by the outward progression of the H-burning shell reaching regions previously homogenized by the convective envelope (during FDU). This reduces the threshold for any extra mixing process to operate \citep[e.g., thermohaline mixing,][]{Charbonnel2007,Cantiello2010}.   
Therefore, the CF mechanism is unlikely to account for \Li{} enhancement observed among main sequence turn-off (MSTO), subgiant, or early RGB stars.
The picture is actually even more complex, since depending on the speed and depth of the extra mixing mechanism, both enrichment \citep[e.g.,][]{Eggleton2008} and depletion \citep[e.g.,][]{Palacios2001,Charbonnel2005,Denissenkov2009} can in principle occur at this evolutionary stage. 

\subsubsection{Mass Transfer Mechanisms}
Accretion processes can lead to changes in stellar abundances. Examples include accretion from novae explosions \citep[e.g.,][]{Martin1994,Izzo2015,Tajitsu2015}, supernova remnants \citep{Woosley1995}, and AGB companions \citep[e.g.,][]{,Kirby2016}. 
However, AGB accretion cannot account for isolated enriched stars, and an enhanced binarity rate is not observed among \Li{}-rich giants \citep{Adamow2018}.

Classical novae can produce Li-rich ejecta \citep{Starrfield1978,Molaro2016}. 
It was estimated that approximately 10\% of \Li{} in the Galaxy is produced by these objects \citep{Rukeya2017}. A significant amount of \Li{} is also produced in Type II supernovae \citep{Dearborn1989}. 
Calculating how much Li-rich ejecta can effectively be accreted onto a star is not straightforward, and it is, therefore, difficult to determine if such channels may account for the observed \Li{}-rich giants.  

\subsubsection{Substellar Accretion or Engulfment}
The accretion or engulfment of a substellar companion---such as a rocky planet, HJ, or brown dwarf---may explain enrichment among isolated stars at early phases (before the LB) of evolution \citep[e.g.,][]{Alexander1967,Siess1999,Villaver2009,Adamow2012}. 
In such a scenario, the companion enhances the host star with its preserved \Li{} supply.
The abundance measurements from CI-chondrites indicate that these rocky bodies are $\sim{}100 \times$ richer in lithium---per unit mass---when compared with the photospheric abundances of their stellar hosts. 
We refer the reader to \cite{Lodders2019} for an in-depth discussion of isotopic abundances from CI-chondrites.

\cite{MacLeod2018} estimated an engulfment occurrence rate of $0.1-1$ event/year in the Galaxy.
Engulfment events may be slightly more common than expected, however, as this estimate does not take into account dynamical effects such as the inward migration arising from Kozai-Lidov oscillations \citep{Frewen2016}. 
The high eccentricities measured among the population of close-in giant planets orbiting evolved stars indicate the strong influence of Kozai-Lidov oscillations \citep{Adamow2018,Grunblatt2018}.

Estimated engulfment occurrence rates also do not account for the presence of binary or triple systems, which comprise ${\sim}80\%$ of A-type stars ($1.4-2.1$~\Msun{}) \citep[e.g.,][]{Kouwenhoven2007,Peter2012}.
\cite{Stephan2018} used computational models to investigate the orbital evolution of HJs in multiple systems, including the effects of the eccentric Kozai-Lidov mechanism (gravitational perturbations induced by a distant stellar companion), general relativity, post-MS stellar evolution, and tidal forces.  
They found that $70\%$ of all close-orbiting planets would eventually be destroyed or engulfed.
In $25\%$ of these cases, the planet was destroyed while the star was in the MS phase of evolution. 
These results encourage further investigation of \Li{} enrichment caused by planetary engulfment events.
Nonetheless, the exoplanet census data indicate that close-orbiting companions, and therefore the accompanying engulfment events, are rare. 
%HJs have been found orbiting a mere $1\%$ of sun-like stars \citep{Winn2015}.
To determine which \Li{}-rich stars might be the descendants of planetary engulfment events, it is critical to discern the strength and survival time of the \Li{} signatures. 
We discuss this further in Section~\ref{sec:Engulfment}.
%%%%%%%%%%%%%%%%%%%%%%%%%%%%%%%%%%%%%%%%%%%%%%%%%%%%%%%%%%%%%%%%%%%%%%%%
%.............................................................
%\needspace{3\baselineskip}
\section{GALAH Lithium Baseline Analysis}
\label{sec:GALAH}
The stellar \ALi{} baseline is a dynamical quantity that is dependent upon on stellar age, mass, chemical composition, and convective envelope depth \citep[e.g.,][]{doNascimento2009,Carlos2016}. 
To determine the statistical significance of the \Li{} enrichment signature arising from planetary engulfment, we calculate an observationally-derived stellar \ALi{} baseline.
Given that engulfment-derived enrichment is a rare occurrence
\citep[e.g.,][and references therein]{Charbonnel2000, Casey2019},
we use abundance measurements to measure the median \ALi{} stellar baseline and associated variance for stars of a particular mass, age, and metallicity.
We employ data from the GALAH DR2 data set \citep{Buder2018}, which provides [Li/Fe] abundance measurements at $R=28,000$ for predominantly nearby stars, across a broad range in evolutionary state. 
In total, there are more than 300,000 stars provided with \Li{} measurements. 
We select the subset of 100,000 of the GALAH stars that have mass measurements available from the catalog of \citet{Sanders2018}.
To avoid the coolest and hottest ranges of stellar models, we restrict our range of stellar surface gravity values from 1.6 $<$ \logg\ $<$ 3.8 dex.
A narrow selection in surface gravity along the giant branch is also an effective narrow range in surface temperature. We work in a narrow range of gravity to minimize the systematic artifacts---inherit systematic offsets in measured \Li{} across evolutionary state---that can be imprinted on abundance measurements due to approximate stellar models.
See \cite{Jofre2019} for a review of systematics in synthetic stellar spectra. 
All abundance measurements are ultimately based on physical synthetic stellar spectra. We minimize variations in the abundances due to artifacts of these spectral models by working in this narrow range in surface gravity. 
Furthermore, we implement a goodness of fit of the spectral model to the data determined from the GALAH pipeline of a reduced $\chi^2$ $<$ 3. 

The abundance measurements from these data have associated uncertainties with them, which are driven by the resolution, signal-to-noise, and stellar parameters of each star (parameterized by effective temperature, surface gravity, and overall metallicity). Some of this uncertainty is captured in the flagging of spectra.
The flags are algorithmic choices made by the pipeline developers of the survey \citep{Buder2018}. Empirical testing of the impact of these flags is useful to examine how different selections impact downstream analyses.
Using the most conservative and recommended flag selections in GALAH, (of flag$\_$cannon = 0 and flag$\_${li}$\_${fe} = 0), reduces our sample to only $\approx$ 500 stars. 
Examining a less restrictive flag selection of flag$\_$li$\_$fe $<$ 4 increases the sample to $\approx$ 50,000 stars. 
However, we see no difference in the overall \Li{} distribution across the evolutionary plane when flags are used or not, thereby leveraging the larger sample of stars without indications that the mean measurements of the observational Li are compromised.
Therefore, we allow for the full sample of $\approx$ 100,000 stars to set our baseline expectation for the mean \Li{} distribution across the evolutionary plane (we do not leverage the abundance of any individual star, in which case flags are more important). 
We examine the baseline \ALi{} abundance, selecting only the subset of ${\sim}10^{4}$ solar-like stars ($\pm 0.11$~dex), and convert the [Li/Fe] from GALAH to an absolute abundance measurement.
Absolute \ALi{} abundances are calculated using the equations below, where the solar abundance is taken as $A(\rm Li)_{\odot}\sim 1.07$~dex \citep{Asplund2009}.

\begin{equation}
    \zeta=10^{([\mathrm{Fe/H}]+[\mathrm{Li/Fe}])}10^{(A(\mathrm{Li})_{\odot}-12)}
\end{equation}
\begin{equation*}
    A(\mathrm{Li})=\log(\zeta)+12
\end{equation*}

We combine the GALAH data with data from Gaia DR2, which offer stellar luminosity and effective temperature estimates for each of the GALAH targets using the Gaia Astrophysical Parameters Inference System (Apsis) \cite{Andrae2018}.\footnote{Visit the link below for more information regarding the Gaia Apsis procedures.\\ \url{https://gea.esac.esa.int/archive/documentation/GDR2/Data_analysis/chap_cu8par/sec_cu8par_process/}}
In Table~\ref{table:ranges}, we list the ranges in stellar mass, effective temperature ($T_{\rm eff}$), luminosity ($L_{\star}$), metallicity ($[\rm Fe/H]$), and \ALi{} for the solar-like subset used in our baseline interpolation. 
In this table, luminosity and temperature values are obtained from Gaia DR2, while all other measurements are provided by the GALAH survey.
The median error for the GALAH \ALi{} measurements was $\pm0.04$~dex.

\begin{table}[tb]
\centering
\caption{The bounds of the parameters contained in our merged Gaia DR2 and GALAH DR2 data set. Luminosity and temperature values are provided by Gaia DR2. All other measurements are obtained by GALAH.}
\begin{tabular}[t]{lcc}
\hline
Parameter & Lower Bound & Upper Bound\\
\hline
$M_{\star}$ & 0.8~\Msun{} & 3.1~\Msun{}\\
$[\rm Fe/H]$ & -0.11~dex & +0.11~dex \\
$T_{\rm eff}$ & 3550 K & 7260 K\\
$L_{\star}$ & 0.14~\Lsun{} & 1960~\Lsun{}\\
\ALi{} & -1.7~dex & +3.7~dex \\
\hline
\label{table:ranges}
\vspace*{0.5mm}  
\end{tabular}
\end{table}%

\begin{figure}[tbp]
\includegraphics[width=0.45\textwidth]{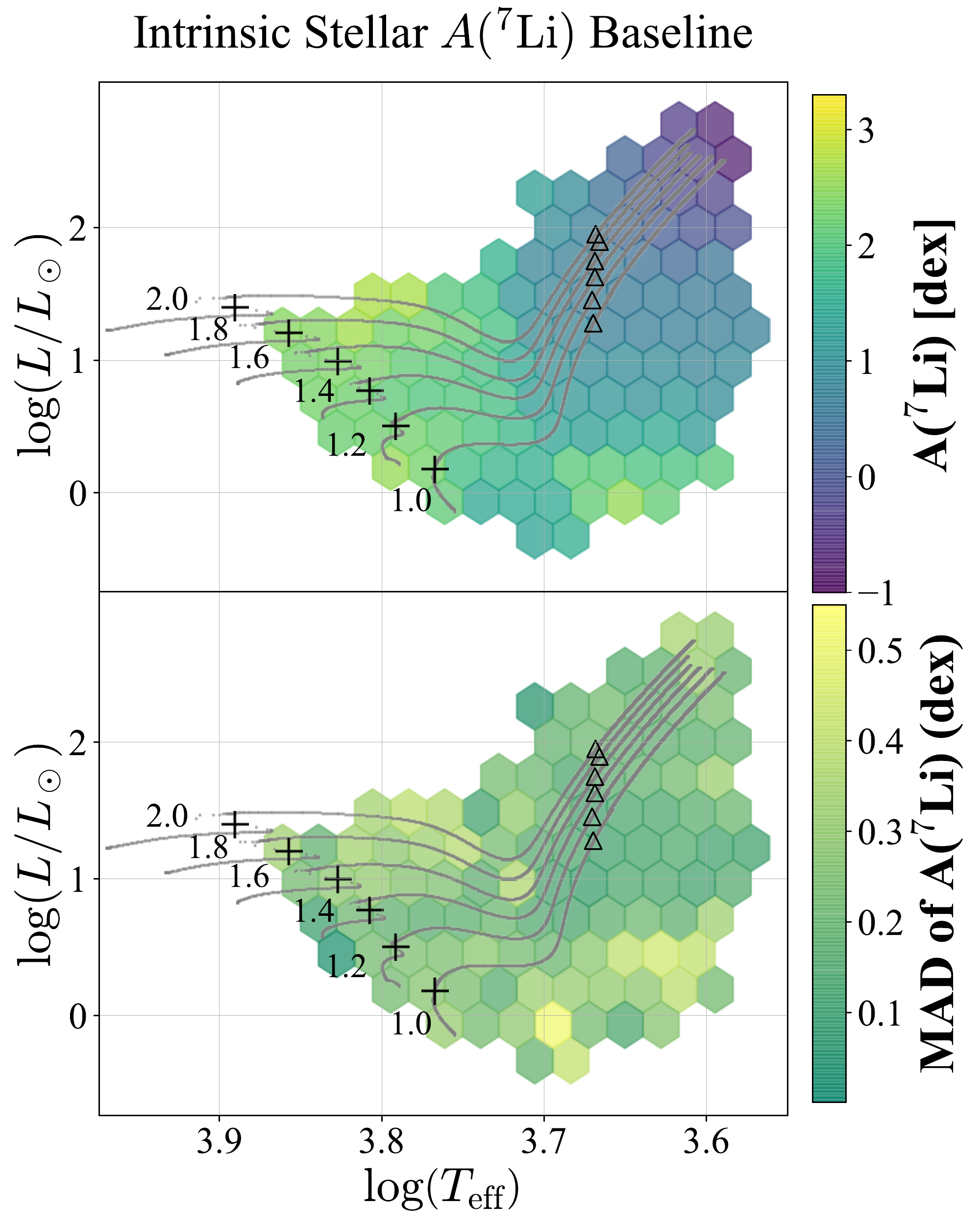}
\caption{
Top Panel: GALAH-derived median \ALi{} measurements for stars from our 2D binned statistic. 
The parameters corresponding to the sample are provided in Table~\ref{table:ranges}.
These values establish our intrinsic stellar \ALi{} baseline. 
As stars evolve, the internal \Li{} supply is diluted.
Increasingly negligible \ALi{} abundances are observed as H-processed material is mixed into the expanding convective envelope. No intrinsic \Li{} enrichment processes were observed among our sample stars.
The light-gray points represent MESA stellar tracks for stars of $1.0-2.0$~\Msun{}. 
The gray plus symbols denote the location of the MSTO, and the end of the FDU episode is denoted by the black triangles.\\
Bottom Panel: GALAH-derived $MAD$ \ALi{} measurements for solar-metallicity stars. 
The $MAD$ range is 0.4~dex across the full parameter space.
The GALAH sample does not cover the early post-MS evolution of a 2.0~\Msun{} star.}
\label{fig:baseline}
\end{figure}

Using the two-dimensional binned statistic from \texttt{SciPy}, the data were binned in two dimensions: $T_{\rm eff}$ and $L_{\star}$.
A total of 225 bins were created ($15\times15$) with a minimum of six stars per bin.
We determined the median \ALi{} abundance and the associated variance for the stars in each bin.
We determined the median absolute deviation from the median ($MAD$) in each bin, which is shown in the bottom panel of Figure~\ref{fig:baseline}.
In the top panel of Figure~\ref{fig:baseline}, we illustrate the data-derived median \ALi{} baseline measurements for solar-metallicity stars in our mass and evolutionary ranges of interest. 
Plotted atop these binned data as light-gray points are the MESA stellar evolutionary tracks for stars ranging between $1-2$~\Msun{}. 
The fact that the GALAH sample does not cover the early evolution of a 2.0~\Msun{} star is not a concern, as the convective zone is far too thin at this stage and is therefore not considered in our investigation. 

To account for the errors associated with the employed parameters, we performed a 1000-sample Monte Carlo simulation, drawing values for $T_{\rm eff}$, \ALi{}, and $L_{\star}$ from a uniform distribution bounded by the corresponding error bars. 
We found that the results are consistent to within $<7\%$ across the plane, with a median error of $1\%$.
Throughout the FDU episode, the measured \ALi{} abundances decrease as H-processed material is mixed into the convective envelope. 
The $MAD$ \ALi{} measurements are shown in the bottom panel of Figure~\ref{fig:baseline}.
The measured variances indicate the thresholds of statistically significant enrichment across this plane, with values ranging between $0.1-0.5$~dex. 
The near homogeneity of the $MAD$ for the \ALi{} across the Luminosity-Temperature plane indicates that there is no remarkable region in evolutionary state in the empirical data where stars are preferentially anomalous. 

%%%%%%%%%%%%%%%%%%%%%%%%%%%%%%%%%%%%%%%%%%%%%%%%%%%%%%%%%%%%%%%%%%%%%%%%%%%%%%%%
\section{Planetary Accretion and Engulfment Events}
\label{sec:Engulfment}
Planetary engulfment events may produce a host of observable effects.
\cite{MacLeod2018} modeled the orbital decay of an engulfed planetary companion arising from drag forces induced in the outer convective envelope.
Treating the deposition of orbital energy as a power source within the convective envelope, they found that such an event could be associated with a luminosity increase up to a factor ${\sim}10^4$ of the initial stellar luminosity.
\cite{Stephan2019b} found that some planet engulfment events can last for centuries or even millennia.

Planetary engulfment could produce the enhanced rotation observed among some giant stars---an anomaly that is unexplained by single star evolution \citep[e.g.,][]{Peterson1983,Siess1999,Livio2002,Massarotti2008,Carlberg2009,Privitera2016a}.
Consistent with the findings of \cite{Qureshi2018}, analysis by \cite{Stephan2019b} determined that MS stars would be rapidly spinning post engulfment.
Among RGB stars, \cite{Stephan2019b} found that engulfment-induced stellar spins could reach or exceed break-up speed.
The rapid rotation of a convective envelope induced by planetary engulfment may result in strong dynamo-generated magnetic fields \citep{Privitera2016b}.
Rapid rotation has also been linked to \Li{} enrichment. \cite{Carlberg2013} analyzed the rotation rates and Li abundances among evolved, single stars of F-, G- and K-type and found that the most rapidly rotating stars also exhibited the highest \ALi{} abundances. 
The strongest correlation between rotation rate and Li abundance was observed among stars of $1.5-2.5$~\Msun{}. 
In Section~\ref{subsec:Model}, we review our stellar evolutionary models, followed by a discussion of the \Li{} contribution from engulfed HJ companions in Section~\ref{subsec:HJLithium}. 
We present our expected \Li{} enrichment signatures from an engulfed HJ companion in Section~\ref{subsec:EnrichmentSignatures}. 
%.............................................................
\subsection{Stellar Models}
\label{subsec:Model}
We computed our models using the open-knowledge 1D stellar evolution software instrument Modules for Experiments in Stellar Astrophysics  \citep[MESA revision 9793][]{Paxton2011,Paxton2013,Paxton2015,Paxton2018}.
We employed these stellar models to track parameters in the convective envelope, such as mass, density, base temperature, and the globally-averaged convective turnover time. 
We did not use these models to calculate the intrinsic stellar \ALi{} abundances, as the theory and implementation of nonstandard \Li{} depletion and enhancement processes are highly uncertain. 

Instead, as described in Section~\ref{sec:GALAH}, we employed observational abundance measurements from GALAH survey data to calculate the stellar \ALi{} baseline as a function of stellar parameters. 
The variance of these data were used to define statistically significant enrichment thresholds. 
This is in contrast to the generally accepted enrichment threshold of \ALi{}~$\geq1.5$~dex, which is the calculated post-dredge-up abundance expectation for Population I stars that formed with meteoritic abundance strengths \citep{Lambert1980,Brown1989,Mallik1999}. 

We ran six non-rotating stellar models with zero-age MS masses of 1.0, 1.2, 1.4, 1.6, 1.8, and 2.0~\Msun{}.
We ran all models with a solar metal abundance of $Z = 0.0169$ \citep{Grevesse1998}.
We ran our models from the pMS until the evolving star expanded to a radius  $R_{\star}\gtrsim0.1$~AU, the outer bound for a HJ companion. 
We employed the Schwarzschild criterion for convective instability, with convective velocities calculated according to the mixing length theory with $\alpha_{\rm MLT} = 2.0$.

We implement a post-processing approach to calculate enrichment strengths arising from HJ engulfment at a given instance in time, which does not account for feedback processes induced by the engulfment of planetary companions. 
These effects could include rotationally-induced mixing, changes in the convective envelope properties, and/or stellar mass loss.
Also not included are effects induced by diffusion, overshooting, and cool bottom processing (e.g., thermohaline mixing).
A beneficial next step would be to model planetary engulfment in a self-consistent manner and to incorporate nonstandard stellar processes. 

Our analysis explores both a simple approach of enrichment, which assumes total dissolution of the engulfed companion in the outer convective region and the ram pressure derived dissolution criterion discussed in \cite{Jia2018}.
We discuss the impact of using different dissolution criteria in Section~\ref{Dissolution}.
%.............................................................
%.............................................................
\subsection{Model Evolution}
\label{subsec:Kipp}

\begin{figure}[tbp]
\includegraphics[width=0.45\textwidth]{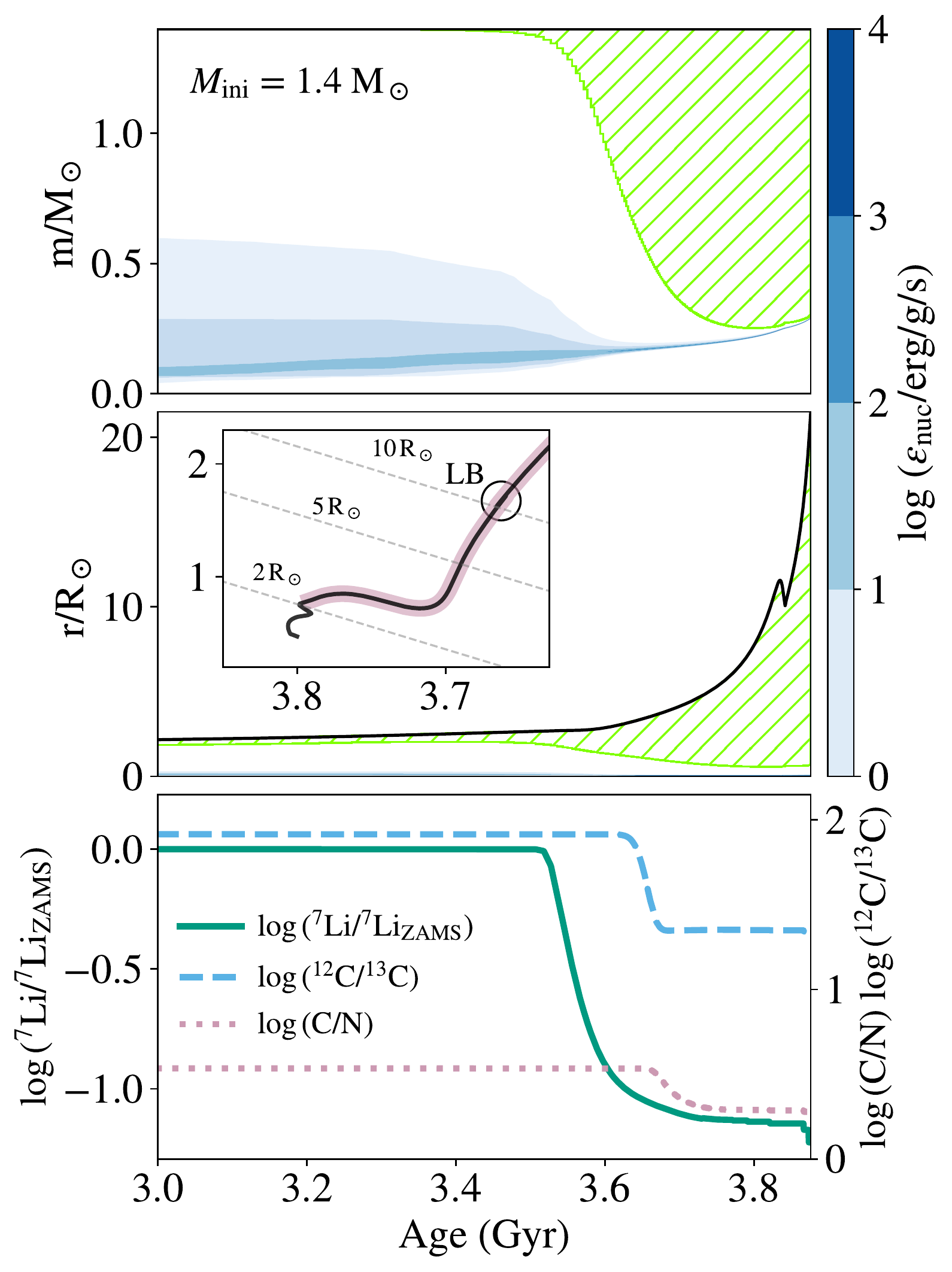}
\caption{Top panel: Kippenhahn diagram showing the evolution of the internal structure of a standard 1.4~\Msun{} model. The Lagrangian mass coordinate is shown as a function of stellar age, from the end of core H-burning and up to the time corresponding to $R_{\star} =0.1$~AU.  Convective regions are hatched green, and regions of nuclear energy generation are shown as blue shading. The first dredge-up starts around 3.55~Gyr. 
Middle panel: same as top panel, but showing the evolution of the radial coordinate. The inset shows the corresponding evolution on the H-R diagram (highlighted in pink). The radial fluctuation at 3.85~Gyr is associated with the LB. 
Bottom panel: time evolution of surface abundance ratios. Lithium decreases very quickly during the first dredge-up, followed by the $^{12}$C/$^{13}$C  and C/N ratios.   
}
\label{fig:Kipp}
\vspace*{2mm}  
\end{figure}

\begin{figure}[tbp]
\includegraphics[width=0.45\textwidth]{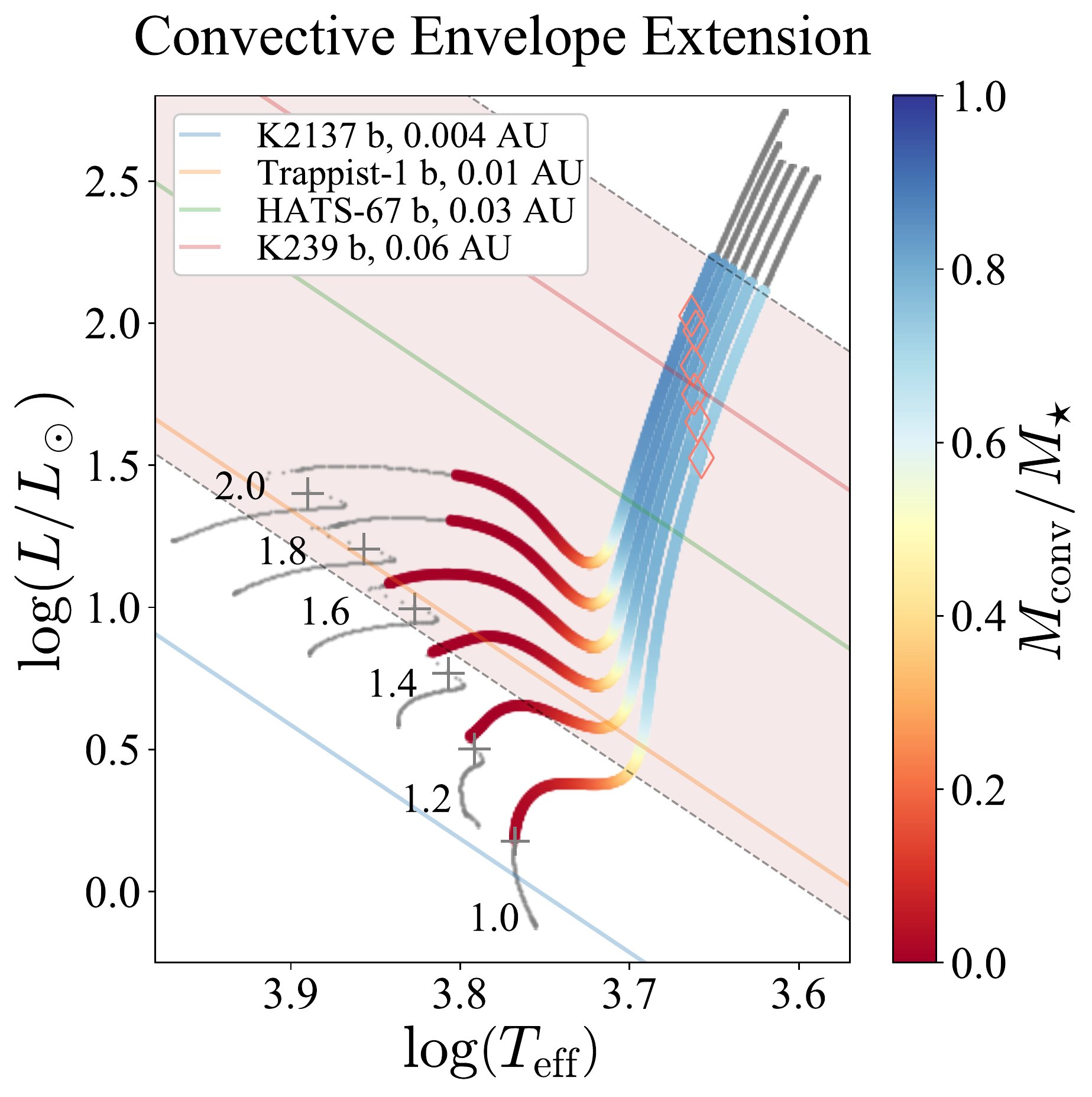}
\caption{Ratio of the mass contained in the outer convective envelope to the total stellar mass. 
Lower mass stars contain more mass in their convective envelopes at the onset of post-MS evolution, as compared to their more massive counterparts.
At early stages of the RGB, however, this ratio approaches 1 for all stellar tracks. 
As a result, the \Li{} present in the convective region becomes increasingly more dilute as the star ages.
The pink-hued band illustrates the orbital separations ranging between $0.01-0.1$~AU.
The location of the MSTO (gray-hued crosses) and LB (salmon-hued diamonds) are illustrated.
}
\label{fig:figure3}
\end{figure}

As the stellar model evolves past the MS, the outer convective region deepens and expands. During the FDU, the inner boundary of the deepening convective envelope overlaps with regions composed of \Li{}-depleted, H-processed material and, as these regions are convectively mixed, the photospheric chemical abundances are altered.
In addition to the depletion of the \Li{} isotope, the FDU episode is known to decrease the ratio of $^{12}\mathrm{C}/^{13}\mathrm{C}$, the ratio of C/N, and Be, as shown in Figure~\ref{fig:Kipp}. 
%, and the \Lisix{} isotope

In Figure~\ref{fig:figure3}, we illustrate a Hertzsprung-Russell (H-R) diagram, showcasing the evolution of the convective envelope mass and temperature for our modeled stars.
We depict the location of the MSTO (gray plus symbols) and the LB (salmon-hued diamonds). 
The figure is color-coded to reveal the increasing convective envelope mass as the star evolves. 
The mass in the convective envelope is initially a small fraction of the total stellar mass.
As the stellar tracks abruptly rise in luminosity at the onset of the RGB, the ratio of the mass in the convective envelope to the total stellar mass increases to $>0.75$ for all stellar tracks. 
Tracking the mass in the convective envelope is critical to determining the expected strength of the planetary engulfment signatures, which is described in detail in Section~\ref{sec:Engulfment}.

However, \Li{} can also be destroyed in the convective envelope of the star. 
The destruction of \Li{} is dependent upon several other stellar parameters, which include the temperature at the base of the convective envelope, as well as the H mass fraction and stellar density in the convective envelope. 
In Section~\ref{sec:Decay} we describe how we can account for the destruction process using our data-driven approach. 
%which is explored further in Section~\ref{sec:Decay}.
%In the bottom panel of Figure~\ref{fig:figure3}, we illustrate the corresponding evolution of the convective base temperature. 
%The plot is color-coded to reveal the ratio of the convective base temperature to the \Li{}-burning temperature threshold ($T=2.5 \times 10^6$~K). 
%At the early phases of post-MS evolution, the ratio is less than one. Thus, \Li{} burning is inefficient, and the isotope is expected to survive in the envelope at these evolutionary phases. 
%At later stages, the convective base temperature rises and crosses the \Li{}-burning threshold; this occurs when $\log({T_{\rm eff}}){\sim}3.7$~K.
%The burning timescale is dependent upon several factors, including the H mass fraction and density. This is discussed further in Section~\ref{sec:Decay} when we present the expected survival times of the \ALi{} enrichment signatures.   
%From the two panels in Figure~\ref{fig:figure3}, it is expected that engulfment-derived \Li{} signatures generated during the late stages of the RGB will be more dilute and short-lived than the signatures produced at earlier phases of post-MS evolution. 
%.............................................................
\subsection{Hot Jupiter Lithium Supply}
\label{subsec:HJLithium}
The composition of a gas giant planet is similar to that of its host star \citep{Demarcus1958}, aside from observed enhancements in some of the heavier elements that comprise ${\sim}1\%$ of the mass fraction \citep{Podolak1974}.
%\footnote{More recently, \cite{Teske2019} found no clear correlation between stellar and planetary (residual) metallicity.}
The outer envelope is predominately composed of H and He.
This envelope surrounds a rocky core that is estimated to be between $10-40$~\MEarth{} \citep{Guillot2009}.
We refer the interested reader to \cite{Fortney2007} for a review of the planetary core mass.
%core mass is dependent upon the assumed equation of state.

Upon engulfment, a fully dissolved 1~\MJup{}  gas giant with a meteoritic \ALi{} abundance would contribute $N_{\mathrm{Li_{p}}}\sim1.7\times 10^{45}$ \Li{} atoms \citep{Montalban2002} and $N_{\mathrm{H_{p}}}\sim8.5\times 10^{53}$ H atoms to the host star. 
%We assume a 0.75 hydrogen mass fraction for the companion.
The \Li{} supply is liberated within this region if the planet is either accreted onto the host stars or completely dissolved in the outer convective envelope.
The enriched \Li{} abundance signature is then given by
\begin{equation}
    A(\rm{Li}_{\mathrm{eng}})~=~\log_{10} \bigg(\frac{N_{\rm{Li_{\star}}}+N_{\rm Li_{p}}}{N_{\rm H_{\star}}+N_{\rm H_{p}}}\bigg) +12~\mathrm{dex}, 
\end{equation}
where the number of stellar \Li{} atoms and H atoms in the convective envelope are given by  $N_{\rm Li_{\star}}$ and $N_{\rm H_{\star}}$, respectively.
We use stellar parameters provided by MESA to determine
\begin{equation}
    N_{\rm H_{\star}}=M_{\mathrm{cz}}X_{\rm cz},
\end{equation}
where $M_{\mathrm{cz}}$ and $X_{\rm cz}$ is the total mass and hydrogen mass fraction in the convective zone, respectively.
Similarly, we use our \ALi{} baseline-derived measurement to determine
\begin{equation}
    N_{\rm Li_{\star}}=N_{\rm H_{\star}}10^{A(Li)_{\star}}10^{-12}.
\end{equation}

%.............................................................
\subsection{Enrichment Signatures}
\label{subsec:EnrichmentSignatures}

\begin{figure}[tbp]
\includegraphics[width=0.45\textwidth]{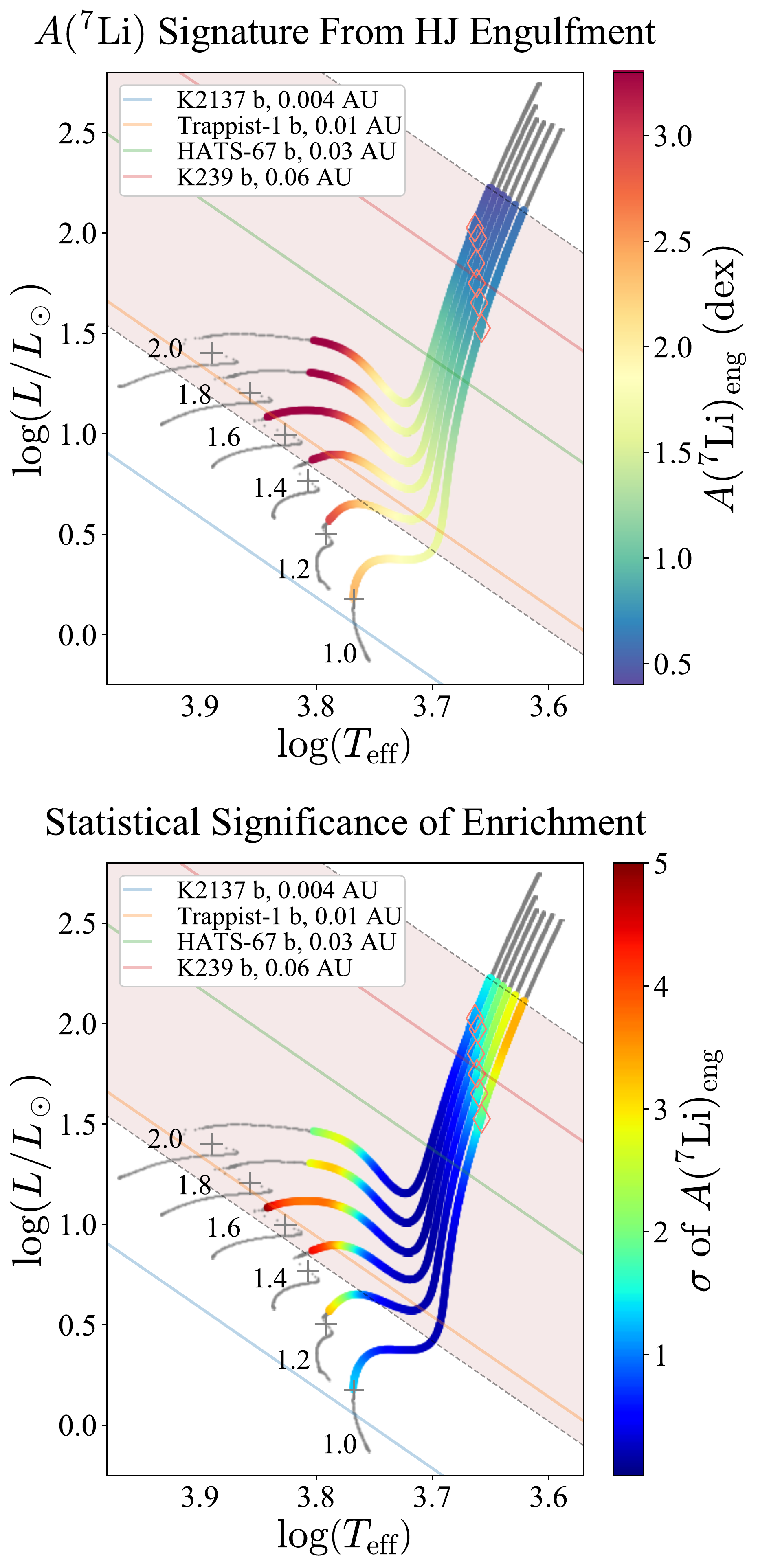}
\caption{Top panel: Estimated enriched \ALi{} abundance measurements for stars that have engulfed a \planetmass{} companion at any point along their evolution.
The location of the MSTO (gray-hued crosses) and the end of the FDU episode (salmon-hued diamonds) are illustrated.
The pink-hued band illustrates the orbital separations ranging between $0.01-0.1$~AU.
The strongest enrichment signatures rival meteoritic abundance strengths (\ALi{}=3.3~dex) and occur in stars of $M_{\star}\gtrsim1.4$~\Msun{} at early phases of post-MS evolution (MSTO to subgiant branch). \\
Bottom panel: The statistical significance of the engulfment-derived enrichment signatures. The most statistically significant signatures (${\sim}5\sigma$) are found among stars of $M_{\star}\gtrsim1.4-1.6$~\Msun{} stars at early phases of post-MS evolution. 
As low mass stars ($\leq 1.2$~\Msun{}) expand to reach $R_{\star}\sim 0.1$~AU, signatures are detectable at a ${\sim}3 \sigma$ confidence level.
} 
\label{fig:figure4}
\end{figure} 
To determine if a companion can produce a significant enrichment signature, we compare the  engulfment-derived \ALi{} enrichment signature for our modeled stars at varying points of stellar evolution to their corresponding stellar \ALi{} baseline. 
We determine the baseline and associated variance for all $T_{\rm eff}$~--~$L_{\star}$ pairs by performing an interpolation routine on the 2d binned data set illustrated in Figure~\ref{fig:baseline}. 
We used a smooth bivariate spline approximation. 

In the top panel of Figure~\ref{fig:figure4}, we illustrate the engulfment-derived \ALi{} enrichment signatures, $A({\mathrm{Li}_{\mathrm{eng}}})$.
When engulfment occurs at the early phases of post-MS evolution, the HJ companion can produce meteoritic abundance measurements (\ALi{}~=~$3.3$~dex).
However, engulfment-derived enrichment signatures are dilute at later stages of post-MS evolution, falling below the generally accepted $1.5$~dex threshold used to designate \ALi{} enrichment in evolved stars. 

In the bottom panel of Figure~\ref{fig:figure4}, we illustrate the statistical significance of the engulfment-derived \ALi{} enrichment signatures, $\sigma_{\rm eng}$.
To determine the statistical significance of the planetary engulfment signature, we compute
\begin{equation}
\sigma_{\rm eng}=\frac{|A({\mathrm{Li}_{\mathrm{eng}}})-A({\mathrm{Li}_\mathrm{base}})|}{MAD},
\end{equation}
where $A({\mathrm{Li}_\mathrm{base}})$ denotes the stellar \ALi{} baseline and $MAD$ is the median absolute deviation from the median of the $A({\mathrm{Li}_\mathrm{base}})$ measurement. 
The $MAD$ values ranged between $0.1-0.5$~dex.

We observe a clear mass dependence at the early phases of post-MS evolution with stronger \ALi{} signatures associated with more massive stars. 
These results rely on the idealized case where either total dissolution of the companion occurs within the outer convective region or the total HJ \Li{} supply is accreted by the host star.
Even with these idealized conditions, statistically significant engulfment-derived enrichment signatures are not observable for 1.0~\Msun{} MSTO and subgiant stars that have engulfed a 1~\MJup{} companion.
These stars have much more mass in their convective envelopes, thereby considerably diluting the \Li{} contribution from the companion.

If one aims to capture statistically significant (${\gtrsim}5\sigma$) enrichment signatures near the MSTO, the best host stars to survey are those between $1.4-1.6$~\Msun{}. 
Above this mass range, the significance drops due to an increase in the $A({\mathrm{Li}_\mathrm{base}})$ and $MAD$ measurements---both increase by ${\sim}0.1$~dex.
When considering early post-MS evolution for the more massive stellar models, there are challenges with an idealized total dissolution assumption.
The shallow convective envelopes among stars of $M_{\star}\geq~1.8~\mathrm{M_{\odot}}$ may not allow for the dissolution and mixing of the companion in this region. 
To help mitigate these concerns, we investigated cases where the stars possess a substantial convective depth, which we chose to be  $\geq 1$~\RJup{}.
It is worth noting that the main  results do not change much by adopting a smaller value for this threshold. 
Moreover, our results apply to the accretion of material from a tidally disrupted substellar companion.
We return to the discussion of the dissolution criterion in Section~\ref{Dissolution}.

The RGB phase is denoted by the sharp rise in luminosity. 
During this phase, engulfment-derived \ALi{} enrichment from a 1~\MJup{} companion is not expected to be statistically significant. 
Moreover, the enrichment strengths expected are well below the Li-rich threshold of 1.5~dex. 
Stars enriched in \Li{} at these stages of evolution cannot be explained by the engulfment or accretion of a HJ companion.
This phase has been associated with intrinsic \Li{} enrichment mechanisms like the CF mechanism.
Our models offer support for the self-enrichment pathway for RGB stars between $1-2$~\Msun{}. 

\begin{figure}[tbp]
\includegraphics[width=0.45\textwidth]{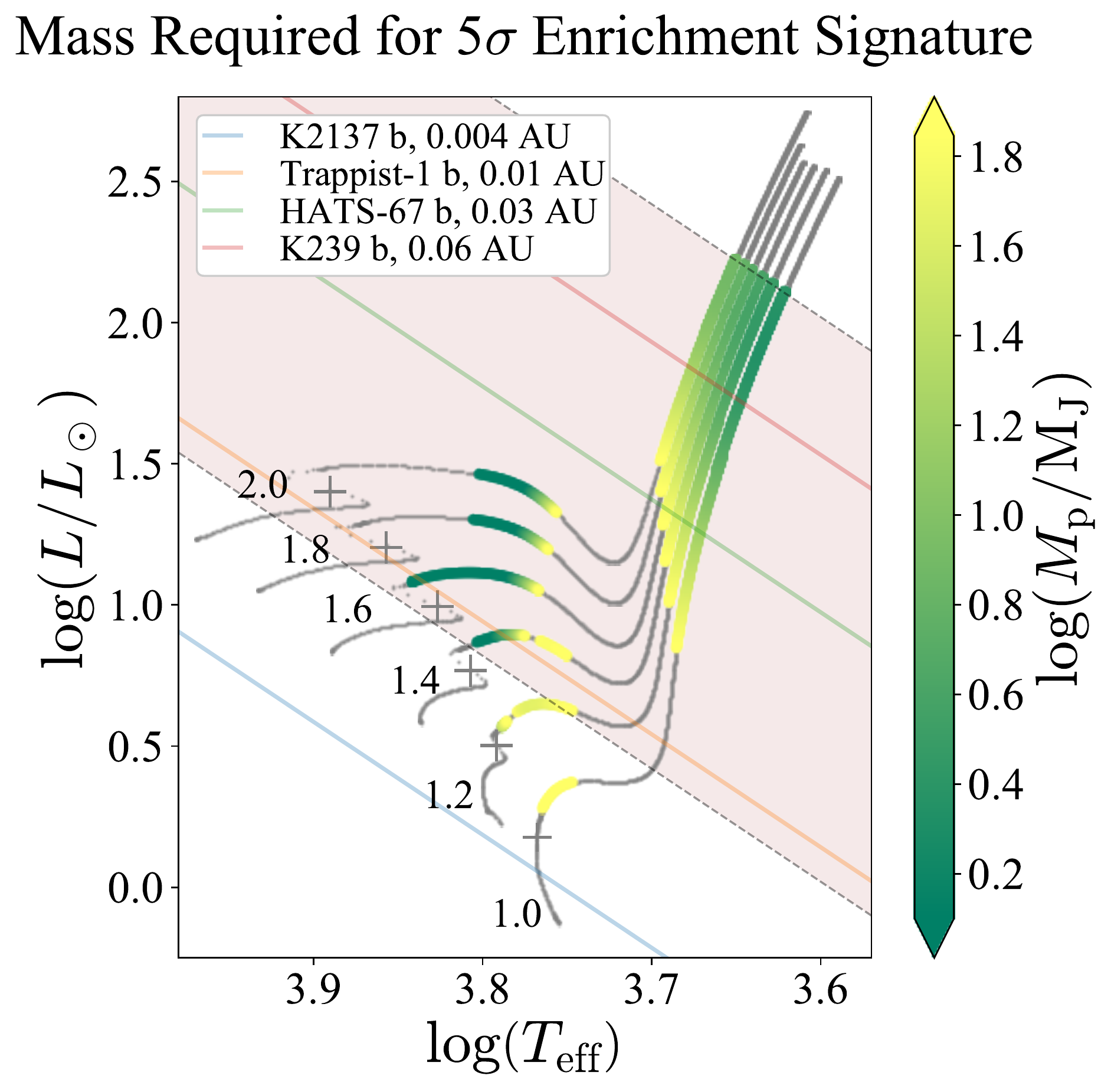}
\caption{Mass of the accreted/engulfed companion required to generate an \ALi{} enrichment signature at a $5\sigma$ confidence level.
Companion masses range between $1-70$\MJup{}. Sources with $M\gtrsim 13$~\MJup{} are classified as brown dwarfs.
Note the parameter space where substellar companions are found to contain insufficient mass to generate a $5\sigma$ \ALi{} enrichment signature (late subgiant and early RG phases).
The pink-hued band illustrates the orbital separations ranging between $0.01-0.1$~AU.
} 
\label{fig:fivesigma}
\end{figure}

In Figure~\ref{fig:fivesigma}, we determine the requisite companion mass to generate an \ALi{} enrichment signature at a $5\sigma$ confidence level. 
We have scaled the companion H and \Li{} supply in this calculation. 
Requisite companion masses $>70$~\MJup{} have been masked, as these sources possess sufficient mass to deplete their \Li{} reservoir. 
Therefore, the illustrated companion masses span between $1-70$~\MJup{}. 
At the MSTO, stars of $1.2-2.0$~\Msun{} are capable of producing $5\sigma$ \ALi{} enrichment signatures with accreted/engulfed companions of $M_{\rm p} \leq 5$~\MJup{}.
For stars on the late subgiant and early RG branches, a single substellar companion does not possess sufficient \Li{} content to generate a $5\sigma$ \ALi{} enrichment signature. 
Enriched stars found in this phase of evolution are therefore likely to be produced by other pathways.

%%%%%%%%%%%%%%%%%%%%%%%%%%%%%%%%%%%%%%%%%%%%%%%%%%%%%%%%%%%%%%%%%%%%%%%%%%%%%%%%
%\needspace{3\baselineskip}
\subsection{Lithium Enrichment Survival Time}
\label{sec:Decay}
%Below I describe the \Li{}-burning prescription outlined in \cite{Andrassy2013}.
To determine the duration of a \Li{} enrichment signature, we employed stellar models to track the surface \Li{} abundance, calculating the time required for the star to return to its former baseline abundance value.
These models also account for some internal mixing processes as well as evolution-induced changes to the stellar parameters.
This includes the convective base temperature and density, which play a major role in \Li{} depletion efficacy, as given by the \Li{}-burning timescale, $\tau_{\rm{Li,burn}}$, 
\begin{equation}
    \tau_{\rm{Li,burn}}=\bigg((9.02) 10^6 X \rho \xi^2 e^{-\xi} \bigg)^{-1} \rm{yr},
\end{equation}
with $\xi$ is given by
\begin{equation}
    \xi= 84.5~T_{6}^{-1/3},
\end{equation} 
and where $T_{6}$ is the convective base temperature in millions of Kelvin, $X$ is the convective envelope hydrogen mass fraction, and $\rho$ is the stellar convective envelope density \citep[see e.g.,][]{Hansen1994,Andrassy2013}.
Also tracked was the time between changes in the abundance baseline, as noted by the GALAH survey data. 
Our models are in good agreement with the depletion timescales provided by \cite{Christensen-Dalsgaard1992}.

\begin{table}[tb]
\centering
\caption{Maximum survival time of surface \Li{} enrichment in a modeled star.}
\begin{tabular}[t]{cc}
\hline
Stellar Mass & Maximum \\ 
& \Li{} Survival Time (Gyr) \\
\hline
1.0~\Msun{} & 2.4 \\
1.2~\Msun{} & 1.5 \\
1.4~\Msun{} & 0.9 \\
1.6~\Msun{} & 0.5 \\
1.8~\Msun{} & 0.2 \\
2.0~\Msun{} & 0.09\\
\hline
\label{table:survival}
\end{tabular}
\end{table}%

\begin{figure}[tbp]
\includegraphics[width=0.45\textwidth]{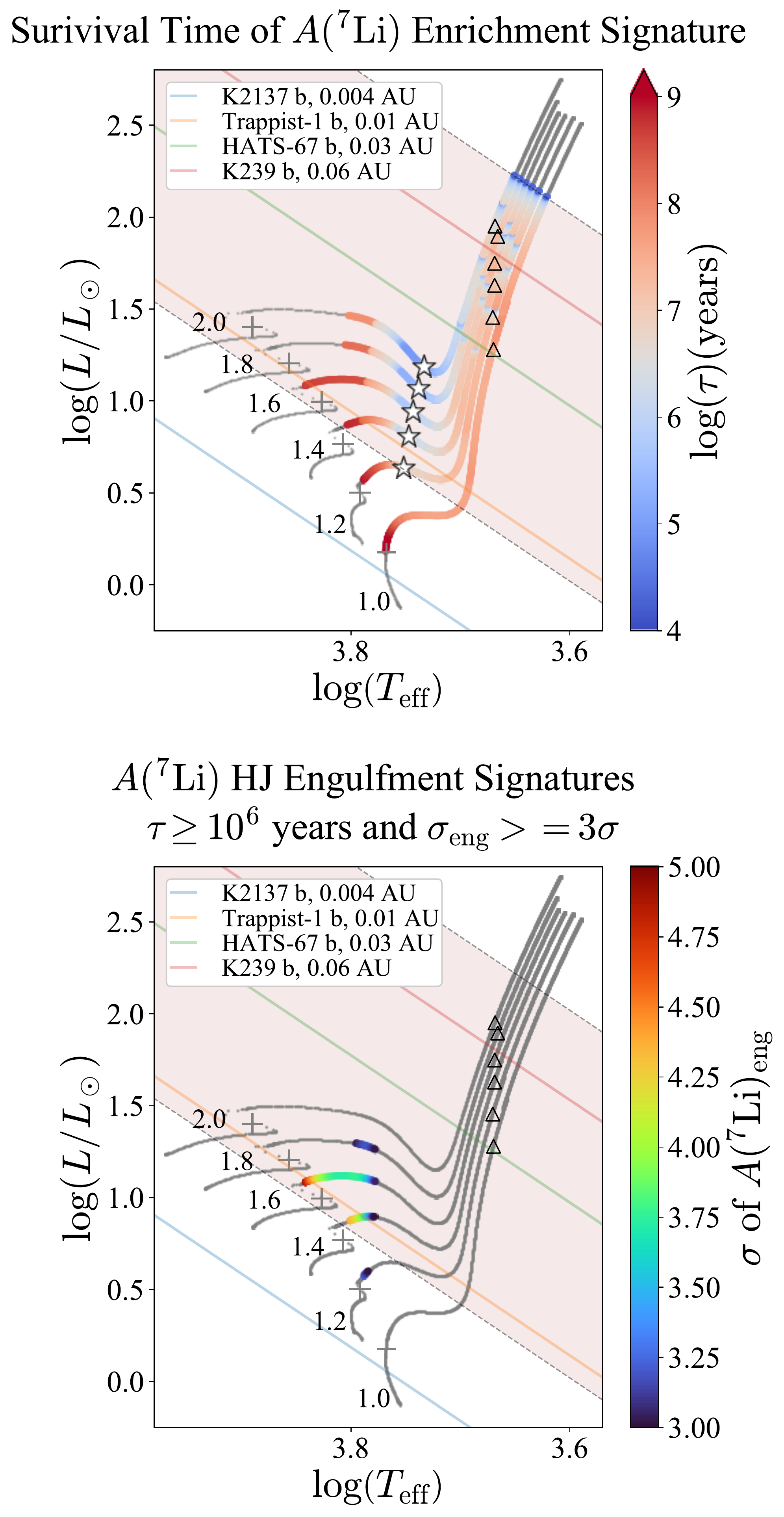}
\caption{Top panel: Survival time of \ALi{} signatures produced from the accretion or engulfment of a \planetmass{} companion.
The white stars denote the point where the convective base temperature reaches the \Li{}-burning threshold.
The pink-hued band illustrates the orbital separations ranging between $0.01-0.1$~AU.
Bottom Panel: Engulfment-derived \ALi{} enrichment abundances for systems where $\sigma \geq 3$~dex and $\tau >= 10^6$~yr.}
\label{fig:figure6}
\end{figure}

To provide a conservative estimate on the \Li{} enrichment depletion time, we compared GALAH survey data and MESA models, taking the smallest of the measurements.
The maximum survival time of the \Li{} supply in the outer convective zone of our modeled stars is provided in Table~\ref{table:survival}.
In addition, we illustrate the survival time of the enriched \ALi{} signatures produced by the engulfment of a \planetmass{} companion in the top panel of Figure~\ref{fig:figure6}.
The white stars denote the location on the HR-diagram where the convective base temperature reaches the \Li{}-burning threshold of $2.5\times10^6$~K.
The \Li{} survival times span a remarkable five orders of magnitude. 
%Our results rule out the possibility of detecting long-lived enrichment signatures for 1.0~\Msun{} stars, as the isotope is rapidly destroyed due to the high convective base temperatures.
%We also rule out the possibility of detecting long-lived enrichment signatures beyond the subgiant branch, including the highly discussed LB. 
%While more massive stars ($M_{\star}\geq 1.6$~\Msun{}) have lower convective base temperatures, their rapid evolution decreases the \Li{} survival time.

We depict our most promising systems in the bottom panel of Figure~\ref{fig:figure6}.
In this figure, we illustrate the engulfment-derived \ALi{} enrichment abundances among systems that meet the following two criteria: the \Li{} is capable of surviving for $\geq 10^6$~yr, and the signature can be observed with a statistical significance $\geq 3\sigma$.
The possibility of detecting enrichment among 1.4~\Msun{} stars near the MSTO is particularly compelling.
Among these systems, the enrichment signatures are expected to survive within the convective envelope for ${\sim}$1~Gyr. 
We predict that spectroscopic surveys will reveal an observational pile-up of \Li{}-enriched stars among ${\sim}1.4$~\Msun{} sources at the early phases of post-MS evolution. 
For stars of $>1.6$~\Msun{} at early post-MS evolutionary phases, \Li{} enrichment signatures are expected to survive for up to $10^8$~yr. 

Our results rule out the possibility of detecting long-lived, statistically significant enrichment signatures for stars evolved beyond the subgiant branch, including the highly discussed LB. 
Enriched stars observed in this region of the H--R parameter space are likely to be produced by a mechanism that continuously replenishes the \Li{} reservoir in the convective region (likely self-generation or the continuous accretion of material from an AGB companion).

\vspace*{8mm}  
%%%%%%%%%%%%%%%%%%%%%%%%%%%%%%%%%%%%%%%%%%%%%%%%%%%%%%%%%%%%%%%%%%%%%%%%%%%%%%%%
\section{Discussion}
\label{sec:Discussion}

\subsection{Dissolution Criterion for Planetary Engulfment Events}
\label{Dissolution}
The results presented in Section~\ref{sec:Engulfment} assume that a \planetmass{} companion has either fully accreted or dissolved within the outer convective zone. 
In the case of engulfment, this assumption is particularly troublesome when one considers the more massive stellar tracks, as these stars harbor thin, tenuous convective envelopes. 
We explore two distinct mechanisms to discern if total planetary dissolution is a reasonable assumption: thermal and mechanical dissolution.

The thermal dissolution criterion relies on the assumption that an engulfed companion will be ablated by the surrounding medium.
The criterion is met when the local stellar temperature exceeds the virial temperature of the companion \citep[see for review][]{Privitera2016b}. 
Using the thermal dissolution criterion, \cite{Aguilera2016} explored a wide range of stellar and companion masses, Li abundances, stellar metallicities, and planetary orbital periods for host stars between $1-2$~\Msun{} and with surface gravity values $<3$~dex. 
They found that a 1~\MJup{} companion could be entirely dissolved within the outer convective envelope of an evolved star, but that total dissolution would not occur in the case of companions with $M_{\mathrm{p}}\gtrsim15$~\MJup{}. 

\citet{Jia2018} asserts that thermal dissolution is an ineffective mechanism for disassociating an engulfed planet.
This is due to the large density ratio between the surrounding stellar material and the planetary surface. 
Instead, they claim that a planet is disassociated when the ram pressure of the stellar flow exceeds the gravitational binding energy of the planet---a process known as \textit{splitting}. 
According to \citet{Jia2018}, a global deformation resulting in planetary dissolution occurs when
\begin{equation}
     f = \frac{\rho_{\mathrm{\star}} v^2}{\rho_{\mathrm{p}} v_{\mathrm{esc}}^2 }>1.
\end{equation}
In this equation, $\rho_{\mathrm{\star}}$ is the density of the surrounding stellar material, $v$ is the orbital velocity of the planet, $\rho_{\mathrm{p}}$ is the density of the planet, and $v_{\mathrm{esc}}$ is the escape velocity of the material at the surface of the planet.
%Global deformation of the engulfed companion occurs when $f>1$.
Our MESA stellar models were augmented to track  $f$ for an engulfed \planetmass{} companion, without accounting for the back-reaction on the stellar structure due to the presence of the engulfed planet.

\begin{figure}[tbp]
\includegraphics[width=0.45\textwidth]{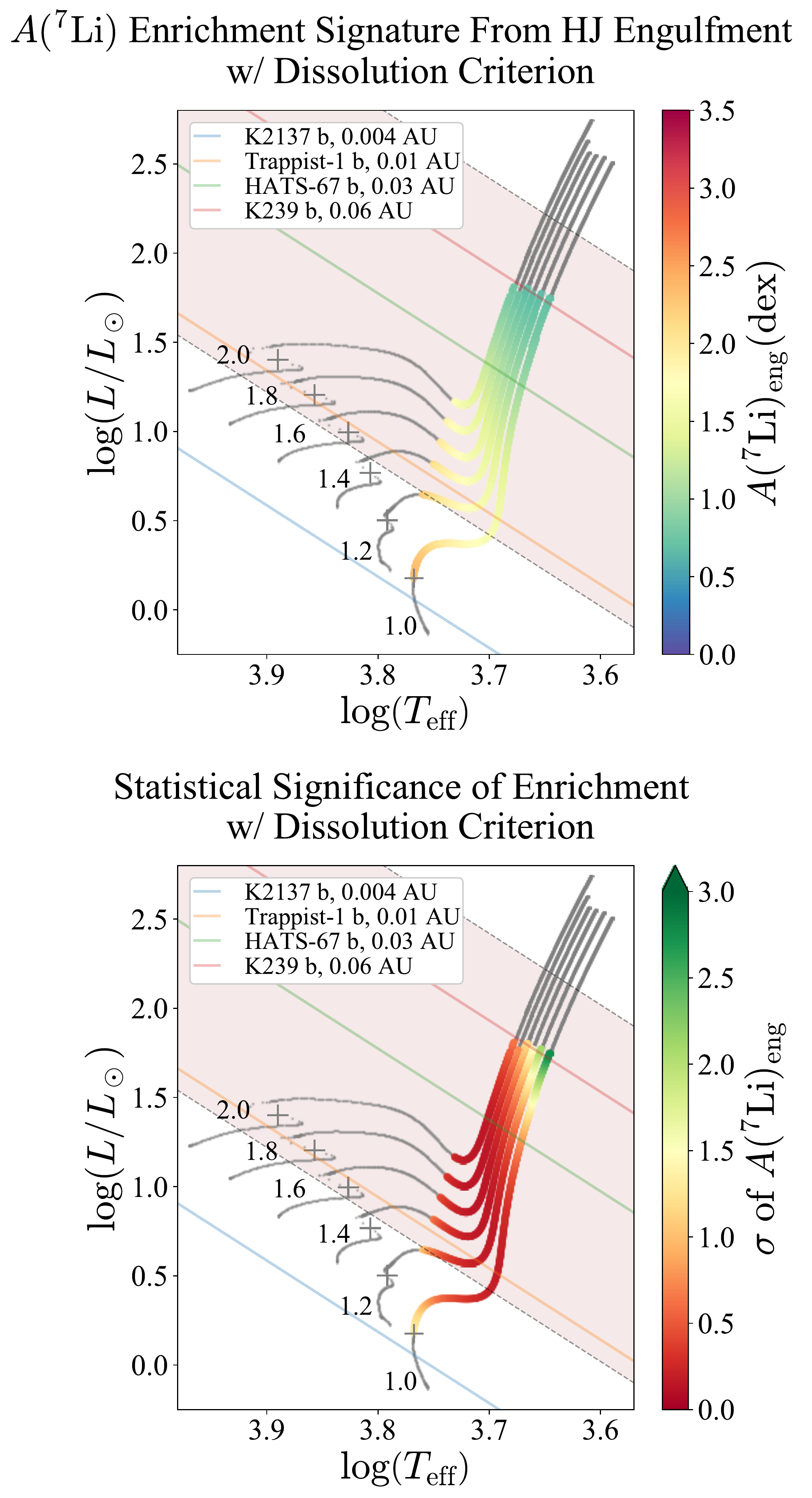}
\caption{Top panel: Application of the global dissolution  criterion from \cite{Jia2018} to \ALi{} enrichment signatures among host stars that have engulfed a \planetmass{ } companion.
The criterion eliminates the statistically significant signatures observed near the MSTO and subgiant branch in Figures~\ref{fig:figure4} and \ref{fig:figure6}.
Moreover, total dissolution is never expected for massive stars of $M_{\star}\gtrsim2.5~\mathrm{M_{\odot}}$.
The pink-hued band illustrates the orbital separations ranging between $0.01-0.1$~AU.
\\
Bottom panel: The statistical significance of the results illustrated in the top panel.}
\vspace*{4mm}  
\label{fig:figure7}
\end{figure}

In Figure~\ref{fig:figure7}, we illustrate the cases where a 1~\MJup{} companion is fully dissolved within the convective region in the two panels. 
The panels are identical to those depicted in Figure~\ref{fig:figure4}, however, we have masked the track points where the global deformation criterion was not satisfied.
In many systems, particularly those at earlier phases of post-MS evolution for stars of $M_\star \geq 1.2$~\Msun{}, the engulfed companion is not expected to completely dissolve in the convective envelope. 
In addition, all scenarios where the companion fully dissolves result in sub-meteoritic enrichment levels ($<3.3$~dex). 
Moreover, a large fraction of these systems do not meet the \ALi{}~$=1.5$~dex enrichment threshold.
If total dissolution of the engulfed \planetmass{} companion is required and if dissolution occurs by the \textit{splitting} process, statistically significant engulfment signatures cannot be produced at a $5{\sigma}$ confidence level. 

Nevertheless, the results discussed in Section~\ref{sec:Engulfment} remain applicable to stars that have partially dissolved or accreted \planetmass{} of material from a substellar companion. 
Moreover, the aforementioned dissolution criteria do not account for the tidal forces generated by the host star, which aids in the dissolution process.
A useful next step would be to perform a self-consistent model of the inspiral and dissolution processes for a companion in close orbit about an evolving star. 

%.............................................................
\subsection{Candidates for Prior Engulfment Events}
\label{Candidates}
The predominately coeval nature of stellar clusters makes these systems an ideal location to search for peculiar enrichment signatures. 
Two \ALi{} enriched giants were found in the Trumpler 20 cluster ($M_{\star}\sim1.8$~\Msun{}) \citep{Smiljanic2016}.
\cite{Aguilera2016b} attributed the \ALi{} enhancement to the engulfment of planetary companions with masses of 15~\MJup{} and 17~\MJup{}.
As discussed in Section~\ref{Dissolution}, dissolving such a companion in the outer convective envelope would be challenging if dissolution relies on the global deformation criterion.
Such a signal would, however, survive for up to $10^8$~yr, as discussed in Section~\ref{sec:Decay}.

The K3III giant star ($2.6$~\Msun{}) BD~+48~740 is an enriched giant (\ALi{}=2.3~dex) with a highly-eccentric 1.6~\MJup{} companion.
Radial velocity variation measurements infer the presence of highly eccentric companions, which are known to destabilize the orbits of other companions. 
Some claim that such systems may be the sites of planetary engulfment events, particularly if the host star displays \ALi{} enrichment.
While we do not probe beyond 2.0~\Msun{}, an extrapolation of our results suggests that engulfment near the MSTO could result in statistically significant enrichment strengths that would survive for over 1~Myr.
The trouble, once again, with such a massive star would be the assumption of total dissolution if global deformation is required to disassociate the companion.

%.............................................................
\subsection{Candidates for Future Engulfment Events}

\begin{figure}[tbp]
\includegraphics[width=0.45\textwidth]{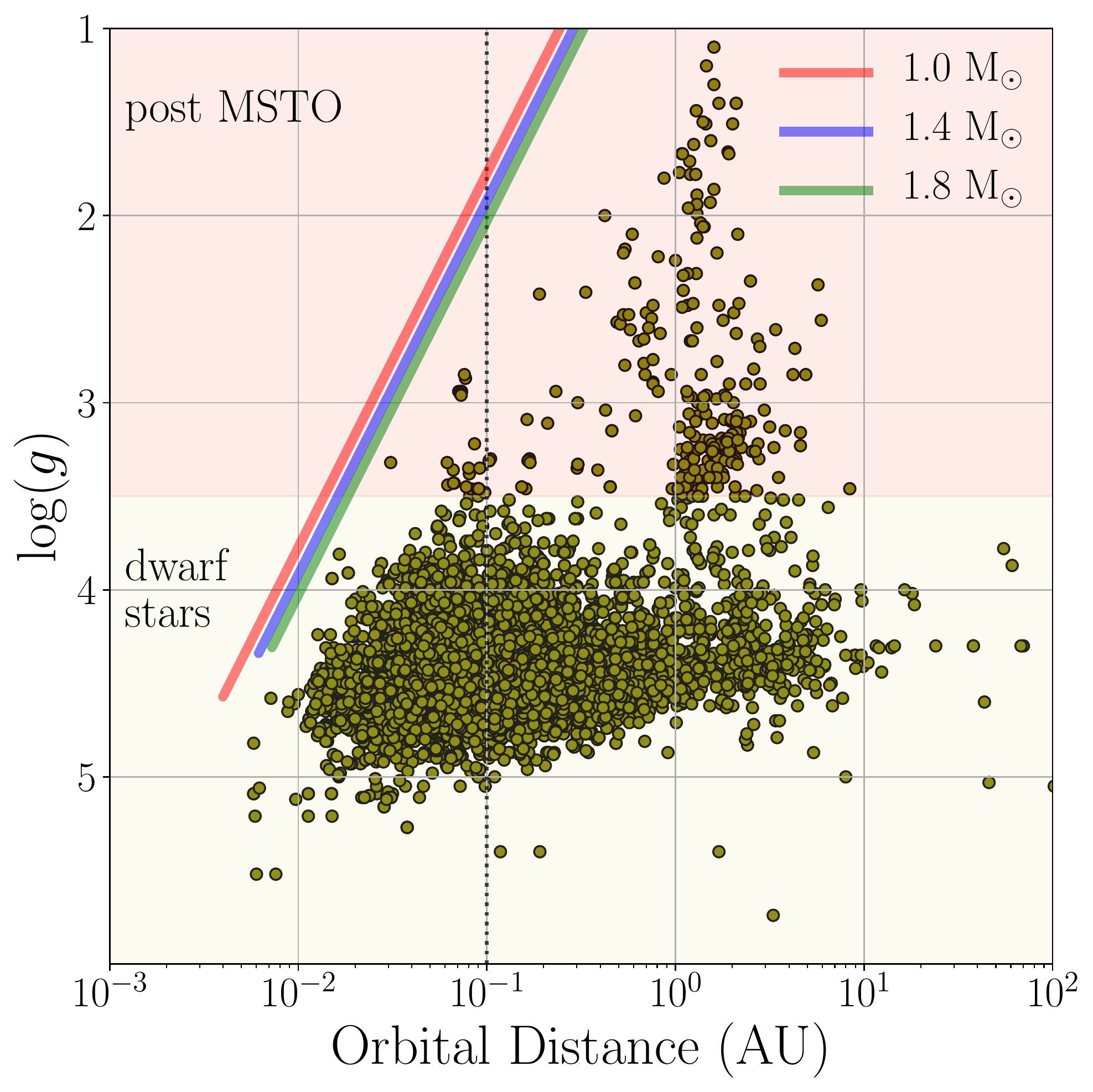}
\caption{Orbital separation and stellar surface gravity for systems with candidate exoplanet sources (taken from \textit{NASA Exoplanet Archive}).
The blue dotted line at 0.1~AU denotes orbital separation cut-off for HJ companions. 
Typical \logg{} values for dwarf stars, subgiants, and giant stars are 4.5, 3.0, and 1.5~dex, respectively. 
Note the dearth of low surface gravity (evolved) stars observed with close-orbiting companions ($a<0.1$~AU). Solid lines show the location in $R_{\star}$-- \logg{} space of stellar evolution models with masses 1.0, 1.4, and 1.8 \Modot{}.
}
\vspace*{4mm}  
\label{fig:planets}
\end{figure}

Few planets have been discovered around evolved stars, particularly in cases where $a\leq 0.5$~AU
\citep[e.g.,][]{Bowler2010,Johnson2010,Villaver2014}.
This is readily observed in Figure~\ref{fig:planets}, where we plot the orbital separation and stellar surface gravity among host stars with exoplanet candidate detections. 
The blue dotted line at 0.1~AU denotes the designated orbital separation cut-off for HJ companions. 
In contrast to the densely populated dwarf star region (\logg{}~$\sim 4.5$~dex), there is an apparent dearth of exoplanet detections among subgiant stars (\logg{}~$\sim~3.0$~dex) and giant stars (\logg{}~$\sim 1.5$~dex). 

However, a lower planet yield is expected among evolved stars due to observational limitations, as detecting planets orbiting evolved massive stars is challenging given an increase in stellar jitter and a diminished transit depth \citep{Delgado2019}. 
Some claim that the scarcity of close-in planets orbiting evolved stars is largely attributable to observational biases, as the majority of surveys targeted subgiants or low-luminosity giants \citep[e.g.,][]{Sato2005,Bowler2010,LilloBox2014,Barclay2015,Jones2016, Grunblatt2016, Grunblatt2017,Nielsen2019}.

Despite these observational hurdles, there are observations of close-in planets orbiting evolved stars.
One remarkable detection is the planet K2 39b, which was observed by \textit{Kepler} and confirmed by the High Accuracy Radial velocity Planet Searcher (HARPS).
The rocky companion is brought within $a/R_{\star}\sim3.4$ of its subgiant host (\logg{}~$\sim 3$~dex) and is expected to survive for only another 150~Myr \citep{VanEylen2016}. 
Similarly, the planet TYC~3663-01966-1~b is expected to be engulfed by its G-type giant host star (\logg{}~$\sim2$~dex), as the companion is brought within $a/R_{\star}\sim 4.5$ of its host star upon periastron \citep{Adamow2018}. 

Data collected by the recent Transiting Exoplanet Survey Satellite (TESS) \citep{Ricker2015} will provide more clarity on the occurrence rates of planets closely orbiting giant stars, as the mission is expected to provide the first statistically significant sample of such systems. 

Recent TESS observations have provided candidate sources within this gap, such as HD 1397b, a giant planet found in an 11.5~day orbit about a G-type subgiant of 1.3~\Msun{} \citep{Brahm2019}. The periastron passage brings this companion within $a/R_{\star}\sim 8$ (0.1 AU) of the host, placing it at risk of disruption via Roche-lobe overflow and tidal inspiralling \citep{Nielsen2019}. 
Such a companion would be ideal to monitor for signs of orbital decay. 
The decay of WASP-12b, a close-orbiting companion ($P{\sim}1$~day) around a late F-type star, was recently observed and interpreted to be the result of tidal dissipation \citep{Yee2020}.
Other TESS relevant detections include a hot Saturn in a 14~day orbit about a late subgiant star \citep{Huber2019} and a 5~\MJup{} companion in a  9~day orbit about a slightly evolved G-type star \citep{Rodriguez2019}.

%%%%%%%%%%%%%%%%%%%%%%%%%%%%%%%%%%%%%%%%%%%%%%%%%%%%%%%%%%%%%%%%%%%%%%%%%%%%%%%%
\section{Summary and Conclusion}
\label{sec:summary}

%main summary and best system identified
Substellar engulfment events have long been suggested as a viable mechanism to explain the \ALi{} enrichment observed among post-MS stars.
We considered the engulfment of close-orbiting ($\lesssim0.1$~AU) \planetmass{} companion among solar-metallicity stars between $1-2$~\Msun{}, using stellar evolutionary models and a data-driven \ALi{} abundance baseline in our analysis. 
Our findings are summarized in the list below.
\begin{itemize}
    \item[--] We determine that the optimal conditions to observe an engulfment-derived enrichment signature are among $1.4-1.6$~\Msun{} stars near the MSTO or at the early phases of the subgiant branch.
    These signatures are capable of surviving in the convective region for up to 1~Gyr. 
    \vspace*{-1mm}  
    \item[--] The engulfment of a \planetmass{} companion is capable of producing meteoritic abundance strengths (${\sim}3.3$~dex) for $\geq 1.4$~\Msun{} stars near the MSTO and at the early phases of the subgiant branch.
     \vspace*{-1mm}  
   \item[--] For stars on the RGB, the engulfment of a \planetmass{} companion cannot result in \ALi{} abundances above the traditionally accepted 1.5~dex threshold. Moreover, these enrichment signatures are not found to be statistically significant.
    \vspace*{-1mm}  
    \item[--] If the total dissolution of the \planetmass{} companion occurs via the splitting mechanism of \citet{Jia2018}, then for many stars the only regions that will be efficiently polluted are below the convective zone. 
    Surface abundances may still be affected if the star accretes the planet via Roche-lobe overflow or if the planet is tidally disassociated before traversing the convective envelope.
\end{itemize}

%future work
It would be valuable to explore the engulfment signatures that arise from other key isotopes, such as \Lisix{} and \Benine{}.
An investigation of \Benine{} is particularly compelling, given that this isotope has a burning temperature that is one million degrees higher than the \Li{} isotope.
Such a signature is therefore measurable at later stages of stellar evolution and could prove to be a useful way of discerning enrichment at the early stages of the RGB. 
This isotope does present some challenges, however, as abundance strengths are weaker than those produced by \Li{} \citep{Reddy2016}.
In addition, observations of the \Benine{} spectral lines are difficult due to severe blending challenges in defining a continuum in this region. Approaches that adopt full spectral fitting and do not rely on conventional continuum normalization may provide a promising avenue for the inference of \Benine{} from spectral data \citep{Ness2015, Ting2019, Leung2019}.

In regard to our approach, a useful next step for future work would be to model planetary engulfment in a self-consistent manner, incorporating non-canonical stellar processes, such as rotationally-induced mixing, diffusion, overshooting, cool bottom processing, or thermohaline mixing, as well as engulfment feedback mechanisms, such as the expansion of the convective envelope and/or stellar mass loss.

% #####################################################################  
\acknowledgments
We gratefully acknowledge helpful discussions with A.~Wheeler, A.~Casey, J.~Auman, E.~Ramirez-Ruiz, A.~Mao, and C.~Malecki. 
This material is based upon work supported by the National Science Foundation Graduate Research Fellowship Program under Grant No. DGE-1656466 and under Grant No. 1909203. Any opinions, findings, conclusions, or recommendations expressed in this material are those of the authors and do not necessarily reflect the views of the National Science Foundation. M.M. acknowledges support for this work provided by NASA through Einstein Postdoctoral Fellowship grant number PF6-170169 awarded by the Chandra X-ray Center, which is operated by the Smithsonian Astrophysical Observatory for NASA under contract NAS8-03060. The Center for Computational Astrophysics at the Flatiron Institute is supported by the Simons Foundation. M.~Ness acknowledges funding from the European Research Council under the European Union's Seventh Framework Programme (FP 7) ERC Advanced Grant Agreement n. [321035].
This work has made use of data from the European Space Agency (ESA) mission {\it Gaia} (\url{https://www.cosmos.esa.int/gaia}), processed by the {\it Gaia} Data Processing and Analysis Consortium (DPAC, \url{https://www.cosmos.esa.int/web/gaia/dpac/consortium}). Funding for the DPAC has been provided by national institutions, in particular, the institutions participating in the {\it Gaia} Multilateral Agreement.  
Lastly, we thank the reviewer for their careful reading of the manuscript and their constructive assessment of our work.

% #####################################################################  
%\vspace{5mm}
%\facilities{HST(STIS), Swift(XRT and UVOT), AAVSO, CTIO:1.3m,
%CTIO:1.5m,CXO}
\facilities{GALAH DR2, Gaia DR2, Exoplanet Archive, ADS}
\software{Astropy\footnote{http://www.astropy.org} \citep{astropy2018}, SciPy \citep{scipy}, MESA \citep{Paxton2011,Paxton2013,Paxton2015,Paxton2018},  Pandas \citep{Pandas}, Matplotlib \citep{matplotlib}, Numpy \citep{numpy}}

% #####################################################################  
%\begin{appendix}

%\section{The 12-point Scale}

%\label{sec:appendixltt}
%Photospheric abundances are calculated on the 12-point scale as $A(\mathrm{X})~=~\log_{10} (N_{\rm \mathrm{X}}/N_{\rm H})~+~12$, where $N_{\rm \mathrm{X}}$ and $N_{\rm H}$ represent the number of atoms of species X and H, respectively. 
%The scale accounts for the fact that the least abundant chemical species are found in ratios of $1:10^{12}$ H atoms. For \ALi{} abundance measurements, the Li I resonance line at 6708.8~\AA{} is generally measured. This include the \ALi{} measurements obtained by the GALAH survey.

%\section{Lithium Depletion in Low Mass Stars}
%\label{sec:appendixlowmass}
%Given the diverse range of interior stellar conditions among MS stars it should come as no surprise that not all MS stars deplete their \Li{} reservoir. 
%For instance, the core temperatures of brown dwarf stars ($M_{\star}~\leq~0.06$~\Modot{}) do not reach the \Li{}-burning threshold. 
%As a result, nearly all their initial \Li{} supply is retained throughout the stellar life cycle \citep[e.g.,][]{Zapatero2002,Lodieu2018}. 
%In contrast, in stars of $0.1$~\Modot{}~$\lesssim M_{\star}\lesssim~0.8$~\Modot{} this fragile isotope is readily destroyed during the MS phase, as the convective envelopes extend to depths where temperatures exceed the \Li{}-burning threshold.

%\end{appendix}
% #####################################################################  
%\bibliographystyle{abbrvnat} 
% or try abbrvnat or unsrtnat or plainnat
%\bibliographystyle{abbrvnat}
\bibliographystyle{aasjournal}
\bibliography{bib.bib}
% #####################################################################  

\end{document}